\documentclass{PoS}
\usepackage{graphicx}
\usepackage{bm}
\usepackage{epsfig}
\usepackage{graphics}

\newcommand{\bmp}{\mathbf p}

\title{Nonrelativistic QCD for coloured scalar fields}

\ShortTitle{Nonrelativistic QCD for coloured scalar fields}

\author{Andre~Hoang and \speaker{Pedro~Ruiz-Femenia}
        \\
        Max-Planck-Institut f\"ur Physik (Werner-Heisenberg-Institut), Germany\\
        E-mail: \email{ahoang@mppmu.mpg.de},
	\email{ruizfeme@mppmu.mpg.de}}

\abstract{We present an effective field theory suitable to describe
a non-relativistic particle-antiparticle pair of heavy
scalars based on the gauge group SU(3).
Its formulation is analogous to that of ``velocity NRQCD" (vNRQCD),
a non-relativistic effective theory for heavy quark pairs.
The matching conditions with scalar QCD and the renormalization group evolution of the effective theory
are discussed. 
The non-relativistic framework proposed here suffices to compute scalar-antiscalar
bound state energies at 
next-to-next-to-leading-logarithmic (NNLL) order and 
next-to-leading-logarithmic (NLL) threshold production
of squarks in $e^+e^-$ and $\gamma\gamma$ collisions, in particular of the lighter 
scalar top quark at a future Linear Collider.
}

\FullConference{International Europhysics Conference on High Energy Physics\\
		 July 21st - 27th 2005\\
		 Lisboa, Portugal}

\begin{document}

\section{Introduction}
Supersymmetry (SUSY) requires the existence of two scalar partners (squarks)
for every quark corresponding to its two helicity states. Many models
of SUSY predict that, due to large mixing, the lightest squark could correspond
to one of the mass eigenstates of the third generation with $m_{\tilde{q}} < 500$ GeV,
thus allowing for the production of stop pairs at a future $e^+e^-$ Linear
Collider operating below 1 TeV. A ``threshold scan'' of the total cross section
line-shape at such facility
will yield precise measurements of the stop mass, lifetime and couplings, 
in close analogy to the
program carried out in threshold studies for the top-antitop 
threshold~\cite{TTbarsim}. 
It is then clear that an accurate description of the shape of the squark production
cross section at energies close to threshold will be mandatory
to put on equal footing the theoretical and experimental uncertainties.

Close to threshold squark pairs are produced with small velocities in the center of
mass frame, $m v^2\equiv \sqrt{s}-2m$, $m$ being the heavy squark mass.
The low-energy QCD dynamics of squarks is based on standard QCD if we assume that the
gluino is not much lighter that the electroweak scale. The multi-gluon exchange of
time-like gluons produces 
singular terms at threshold $\propto (\alpha_s/v)^n$ and $\propto (\alpha_s \ln v)^n$  
at the $n$-loop order in QCD perturbation theory, which 
forces one to carry out a double expansion in $\alpha_s$ and
$v$ and count $v\sim\alpha_s$ to define the power counting in this regime. The singular
terms mentioned above must be thus termed as leading order contributions and need to be
summed up to all orders in $\alpha_s$. For top-antitop
pair production at threshold, vNRQCD ("velocity NRQCD'')~\cite{LMR,HoangStewartultra}
has been shown to be
the proper framework to consistently perform the summation to 
NNLL order~\cite{hmst}, which means
to account up to ${\cal O}(\alpha_s^2,\alpha_s v,v^2)$ corrections to the
leading-logarithmic (LL) order $\propto\sum_{k}(\alpha_s/v)^k\sum_{i}(\alpha_s \ln v)^i$.
An analogous effective field theory 
describing the non-relativistic interaction between pairs of colored scalars
can provide the needed ingredients for a renormalization group improved computation
of squark pair production at threshold. A complete summation at
NLL order requires the knowledge of the 
Wilson coefficients of potentials at LL order 
and of nonrelativistic production currents at NLL order~\cite{LMR,HoangStewartultra}.  
The effective Lagrangian for the scalar version of
vNRQCD (that will be called simply vNRQCD in the following) and the 
relevant matching conditions and anomalous dimensions are
outlined in this letter. We refer the reader to a recent
work~\cite{ours} for a more detailed analysis and explicit formulae.

\section{Effective theory set-up}

\begin{figure}
  \epsfxsize=9cm \centerline{\epsfbox{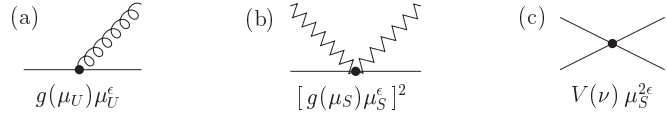}}
  \vskip -0.5cm
  \medskip
{\caption{Effective theory diagrams for the interaction of
a heavy squark with an ultrasoft gluon.}
\label{eftinterac}}
\end{figure}

The vNRQCD effective theory is formulated by including only those quark and 
gluonic degrees of freedom which can become on-shell for energies below $m$.
The on-shell degrees of freedom correspond to
gluons ($A_q^{\mu},\,A^{\mu}$) and ghosts ($c_q,\,c$)
with soft $\sim(mv,mv)$ and ultrasoft $\sim(mv^2,mv^2)$,
energy and momenta, 
and potential heavy squarks ($\psi_\bmp$) and
antisquarks ($\chi_\bmp$), with energy $\sim mv^2$ and soft
three momentum $\bmp\sim mv$. 
Massless quarks and squarks can also be included with soft and ultrasoft 
components in vNRQCD. 
The dependences on soft
momenta of the heavy squark and soft gluons appear as labels 
on the fields and an explicit coordinate dependence refers only to ultrasoft
fluctuations~\cite{LMR}. All off-shell effects such as those from hard squarks
and gluons, potential gluons, and soft quarks are accounted for by
on-shell matching of vNRQCD to full QCD at the hard scale $m$.
The basic vNRQCD Lagrangian can be separated into ultrasoft, soft and potential
components ${\cal L}={\cal L}_u+{\cal L}_s+{\cal L}_p$.
The ultrasoft
piece of the effective Lagrangian describes the interactions of ultrasoft gluons with
squarks (Fig.~\ref{eftinterac})a, which are multipole expanded to separate
ultrasoft momenta from larger momenta, 
and also contains the kinetic energy contributions for the
squarks and the ultrasoft fields:
\begin{eqnarray} 
\label{Lus}
{\cal L}_u  &=& 
\sum_{\mathbf p}\,\bigg\{
   \psi_{\bmp}^*   \bigg[ i D^0 - \frac {\left({\bf p}-i{\bf D}\right)^2}
   {2 m} +\frac{{\mathbf p}^4}{8m^3} + \ldots \bigg] \psi_{\bmp}
 + (\psi \to \chi,T\to \bar T)\,\bigg\}
 -\,\frac{1}{4}G_u^{\mu\nu}G^u_{\mu \nu}  \,,
\label{Lu}
\end{eqnarray}
where $G^{\mu\nu}$ is the ultrasoft field strength tensor and the covariant
derivative, $D^{\mu}=\partial^{\mu}+i\mu_U^{\epsilon}g_u A^{\mu}$, 
only involves the ultrasoft gluon field and the ultrasoft gauge coupling $g_u=g_u(\mu_U)$.
The factor of the ultrasoft renormalization scale $\mu_U$ that appears in the
covariant derivative is determined by the requirement that the kinetic 
terms in the vNRQCD action are of order $v^0$~\cite{LMR,HoangStewartultra}
in $d=4-2\epsilon$ dimensions,
and we employ the same power counting formalism as 
in the heavy quark case~\cite{LMR,HoangStewartultra}. 
Likewise, the velocity scaling
determines the factors of the soft renormalization scale $\mu_S$ accompanying the 
soft gauge coupling $g_s=g_s(\mu_S)$ 
in the soft and potential interactions (Figs.~\ref{eftinterac}b,c). The existence of two different 
subtraction scales
guarantees that each of loop in the EFT receives the 
proper renormalization scale 
according to the three-momenta flowing through it. A crucial feature
of vNRQCD is that $\mu_S$ and $\mu_U$ are required to be correlated according to the 
non-relativistic energy-momentum relation of a heavy particle, i.e.
$\mu_U=\mu_S^2/m=m\nu^2$, where $\nu$ is the dimensionless renormalization parameter
of vNRQCD. Lowering $\nu$ from the hard matching
scale ($\nu=1$) to a scale of order the squark velocity 
sums all logarithms of the soft 
and ultrasoft scales  
into the Wilson coefficients of the vNRQCD operators,
and renders vNRQCD matrix elements small.

\begin{figure}[!t]
  \epsfxsize=8.5cm \epsfbox{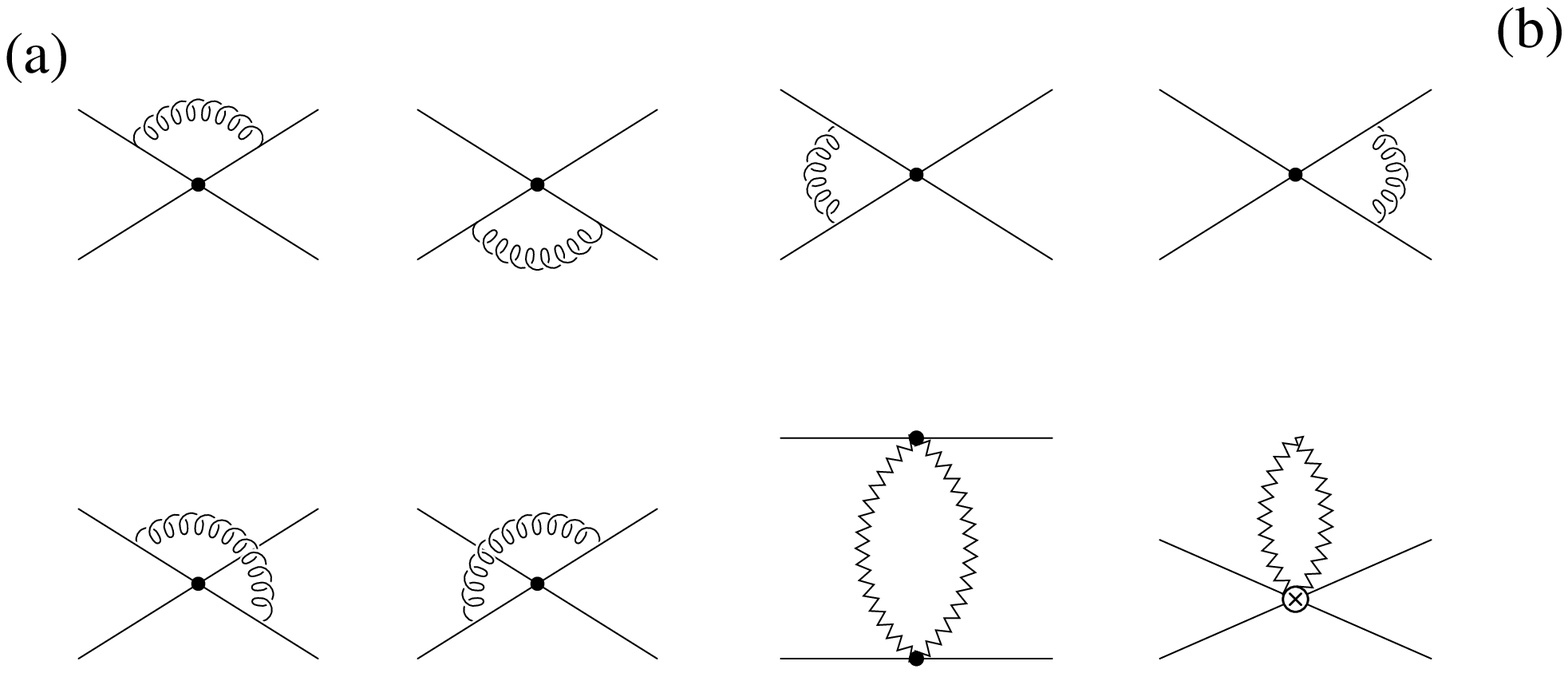}
  \vskip -4.8cm
  \hskip 6.7cm
  \epsfxsize=9.5cm \epsfbox{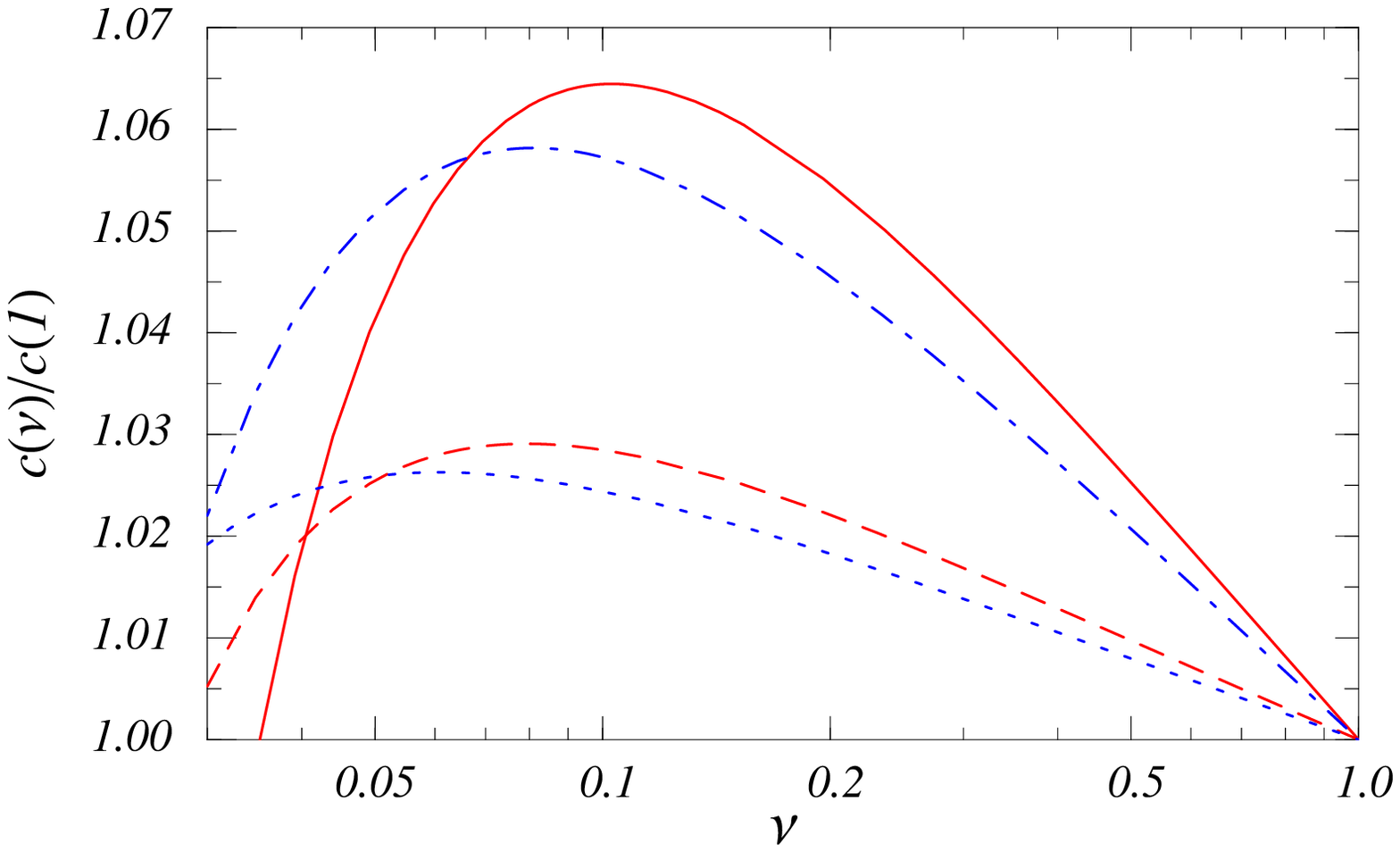}
  \vskip -7.3cm
  \medskip
{\caption{(a) Diagrams for the computation of the LL anomalous dimension of potentials.
(b) NLL running of the normalized Wilson coefficient for S-wave
    and P-wave currents for $m=220$~GeV (solid and dashed lines) and
    for $m=500$~GeV (dot-dashed and dotted lines).}
\label{1loop}}
\end{figure}

The vNRQCD Lagrangian also contain operators that describe interactions of quarks with soft
gluons ${\cal L}_s$ (Fig.~\ref{eftinterac}b), that arise when soft quarks are integrated out, 
and potential-like squark-antisquark interactions ${\cal L}_p$
(Fig.~\ref{eftinterac}c),
originating from potential gluons and other off-shell modes. 
Setting aside spin-dependent contributions that are absent for scalars, 
the matching coefficients of ${\cal L}_s$ in the scalar theory agree 
with the corresponding ones in the
quark theory due to the equivalence of the heavy quark and heavy scalar effective Lagrangians 
(HQET and HSET)
at order $1/m$. Differences arise at NNLO because the Darwin term vanishes at the hard scale in
HSET~\cite{ours}. The coefficients in $ {\cal L}_s$ are only renormalized due to soft 
interactions~\cite{HoangStewartultra,amis} and the structure of the resulting soft
divergences is identical to the one in HSET.
The potential Lagrangian has terms
\begin{eqnarray}
{\cal L}_p &=& -\mu_S^{2\epsilon}\sum_{\bmp,\bmp^\prime} V({\bmp,\bmp^\prime})\,
   \psi_{\bmp^\prime}^* \psi_{\bmp}
   \chi_{-\bmp^\prime}^* \chi_{-\bmp}
 + \ldots \quad,\quad 
V({\bmp},{\bmp^\prime})  =   (T^A \otimes \bar T^A) \bigg[
 \frac{{\cal V}_c^{(T)}}{(\bmp^\prime-\bmp)^2}
 + \ldots \bigg]\,.
\label{Lp}
\end{eqnarray}  
The coefficients ${\cal V}^{(1,T)}$ are obtained by matching to the full theory 
Born scattering diagrams (e.g. for the Coulomb potential shown above, 
${\cal V}_c^{(T)}(1)=4\pi\alpha_s(m)$), and they differ in general
from the corresponding results in the quark case. The LL renormalization
of the potentials are determined by the effective theory one-loop graphs 
shown in Fig.~\ref{1loop}a. Since to the order we are interested in interactions with 
ultrasoft gluons are not sensitive to spin, the 1-loop ultrasoft 
renormalization is the same in vNRQCD for both quarks and squarks. There are
a number of operators in the potential and soft parts of vNRQCD with vanishing
matching conditions at $\nu=1$ which however become non-zero for $\nu<1$ due
to ultrasoft renormalization~\cite{HoangStewartultra}.

\section{Anomalous dimensions for production currents}

At leading order in the non-relativistic expansion heavy squark-antisquarks pairs
are produced in S- and P-wave states in $\gamma\gamma$ and $e^+e^-$
collisions by the currents $J_{S,\bf p} = 
    \psi_{\bmp}^*\,\chi_{-\bmp}^*$ and ${\bf J}_{P,\bf p}=   \psi_{\bmp}^*\,\bmp\,\chi_{-\bmp}^*$
     respectively, with 
    corresponding Wilson coefficients $c_s(\nu)$ and $c_P(\nu)$.
The total cross sections $\sigma(e^+e^-,\gamma\gamma\to\tilde{q}\bar{\tilde{q}})$ 
can be written in terms of the imaginary part of the correlators built from the non-relativistic currents
and the products of their Wilson coefficients. The NLL running of the Wilson coefficients
 is determined by the two-loop
anomalous dimensions of the currents, as calculated in Ref.~\cite{ours}. 
The results are displayed in Fig.\ref{1loop}b. 
The scale
variation for the S-wave coefficient is found to be much stronger than for the P-wave coefficient
and the maxima slightly decrease with the heavy scalar mass and also move towards smaller values 
of $\nu$. 
The remaining scale dependence in the NLL cross section comes from the non-relativistic 
correlators that are proportional to the zero-distance Green's functions 
obtained by solving the Schr\"odinger equation with the NLL improved Coulomb potential.



\begin{thebibliography}{99}

\bibitem{TTbarsim}
M.~Martinez and R.~Miquel,
Eur.\ Phys.\ J.\ C {\bf 27}, 49 (2003);
A.~H.~Hoang {\it et al.},
in Eur.\ Phys.\ J.\ direct {\bf C3}, 1 (2000);
A.~H.~Hoang,
arXiv:hep-ph/0204299.


\bibitem{LMR} 
M.~Luke, A.~Manohar and I.~Rothstein,
Phys.\ Rev.\  {\bf D 61}, 074025 (2000);
 A.V.~Manohar and I.W.~Stewart,
Phys.\ Rev.\  {\bf D 62}, 074015 (2000).

\bibitem{HoangStewartultra}
A.~H.~Hoang and I.~W.~Stewart,
Phys.\ Rev.\ D {\bf 67}, 114020 (2003).

\bibitem{hmst}
A.~H.~Hoang, A.~V.~Manohar, I.~W.~Stewart and T.~Teubner,
Phys.\ Rev.\ Lett.\  {\bf 86}, 1951 (2001);
A.~H.~Hoang, A.~V.~Manohar, I.~W.~Stewart and T.~Teubner,
Phys.\ Rev.\ D {\bf 65}, 014014 (2002);
A.~H.~Hoang,
Phys.\ Rev.\ D {\bf 69}, 034009 (2004).

\bibitem{ours}
  A.~H.~Hoang and P.~Ruiz-Femenia,
  arXiv:hep-ph/0511102.

\bibitem{amis} A.V.~Manohar and I.W.~Stewart,
Phys.\ Rev.\ D {\bf 62}, 014033 (2000)
[arXiv:hep-ph/9912226].


\end{thebibliography}
\end{document}